\documentclass[prl,twoside,twocolumn,superscriptaddress,showpacs]{revtex4}
\usepackage{graphicx}
\usepackage{amsmath}
\usepackage{amssymb}

\begin{document}

\title{Charge order in LuFe$_2$O$_4$: antiferroelectric ground state and coupling to magnetism}

\author{M. Angst}
 \email{m.angst@fz-juelich.de}
\affiliation{Materials Science and Technology Division, Oak Ridge
National Laboratory, Oak Ridge, TN 37831, USA}
\affiliation{Institut f\"ur Festk\"orperforschung,
Forschungszentrum J\"ulich GmbH, D-52425 J\"ulich, Germany}
\author{R.~P. Hermann}
\affiliation{Institut f\"ur Festk\"orperforschung,
Forschungszentrum J\"ulich GmbH, D-52425 J\"ulich, Germany}
\affiliation{Department of Physics, B5, Universit\'e de Li\`ege,
B-4000 Sart-Tilman, Belgium}
\author{A.~D. Christianson}
\author{M.~D. Lumsden}
\affiliation{Neutron Scattering Science Division, Oak Ridge
National Laboratory, Oak Ridge, TN 37831, USA}
\author{C.~Lee}
\author{M.-H.~Whangbo}
\affiliation{Department of Chemistry, North Carolina State
University, Raleigh, NC 27695, USA}
\author{J.-W. Kim}
\author{P.~J. Ryan}
\affiliation{Ames Laboratory, Ames, IA 50010, USA}
\author{S.~E. Nagler}
\affiliation{Neutron Scattering Science Division, Oak Ridge
National Laboratory, Oak Ridge, TN 37831, USA}
\author{W. Tian}
\affiliation{Materials Science and Technology Division, Oak Ridge
National Laboratory, Oak Ridge, TN 37831, USA} \affiliation{Ames
Laboratory, Ames, IA 50010, USA}
\author{R. Jin}
\author{B.~C. Sales}
\author{D. Mandrus}
\affiliation{Materials Science and Technology Division, Oak Ridge
National Laboratory, Oak Ridge, TN 37831, USA}

\begin{abstract}
X-ray scattering by multiferroic LuFe$_2$O$_4$ is reported. Below
$320\,{\rm K}$, superstructure reflections indicate an
incommensurate charge order with propagation close to
($\frac{1}{3}\frac{1}{3}\frac{3}{2}$). The corresponding charge
configuration, also found by electronic structure calculations as
most stable, contains polar Fe/O double-layers with {\em
anti}ferroelectric stacking. Diffuse scattering at $360\,{\rm K}$,
with ($\frac{1}{3}\frac{1}{3}0$) propagation, indicates
ferroelectric short-range correlations between neighboring
double-layers. The temperature dependence of the incommensuration
indicates that charge order and magnetism are coupled.
\end{abstract}

\date{\today}


\maketitle

Materials where ferroelectricity or dielectric behavior is coupled
to magnetism have the potential for novel applications and
presently receive a lot of attention \cite{EerensteinCheong}. A
new type of ferroelectricity, originating from charge order (CO)
and seemingly coupled to magnetism, has been proposed to occur in
LuFe$_2$O$_4$ containing triangular Fe/O double-layers
\cite{Ikeda05}, generating a lot of interest in this material
\cite{Subramanian06,NaganoNaka,sPark07,Xiang07,sLi08b,Zhang07}.
Ferroelectricity is thought to arise from a particular arrangement
of Fe$^{2+}$ and Fe$^{3+}$ within the Fe/O double-layers, making
these intrinsically polar. However, different reported
\cite{Zhang07,Yamada0097} superstructure reflections indicate an
incomplete understanding of the CO, and the full three-dimensional
(3D) charge configuration has yet to be established. The latter
determines the stacking of the polarizations of the individual
double-layers, and thus the net polarization of the material.

Further, although magnetism in LuFe$_2$O$_4$ occurs in the CO
state to date there is no direct observation of coupling between
CO and magnetism to date. Such an observation would be important
in establishing the mechanism by which ferroelectricity,
dielectric behavior, and magnetism couple
via the underlying CO. We have recently grown LuFe$_2$O$_4$ 
crystals with magnetic transitions of unprecedented sharpness, on
which neutron scattering allowed the first refinement of a 3D spin
structure \cite{Christianson08}.

Here, we present a study of the CO superstructure by synchrotron
x-ray scattering. We propose a commensurate approximation for
three domains of the incommensurate CO configuration, with
propagations (in hexagonal notation) close to the
symmetry-equivalent directions
($\frac{1}{3}\frac{1}{3}\frac{3}{2}$),
($\overline{\frac{2}{3}}\frac{1}{3}\frac{3}{2}$), and
($\frac{1}{3}\overline{\frac{2}{3}}\frac{3}{2}$), corresponding to
an {\em anti}ferroelectric \cite{noteAFE3} ground state that we
also identify by first-principles density-functional theory (DFT)
as having the lowest energy. In contrast, short-range charge
correlations above the CO temperature were found to correspond to
a ferroelectric CO configuration with ($\frac{1}{3}\frac{1}{3}0$)
propagation. Further, we provide evidence for a coupling between
CO and magnetism involving primarily the incommensuration of the
CO. Our results underline the importance of near degeneracy of the
CO in LuFe$_2$O$_4$ and provide an essential microscopic basis for
magnetoelectric coupling previously proposed
\cite{Ikeda05,Subramanian06,NaganoNaka,sPark07}.

\begin{figure}[!t]
\includegraphics[width=0.99\linewidth]{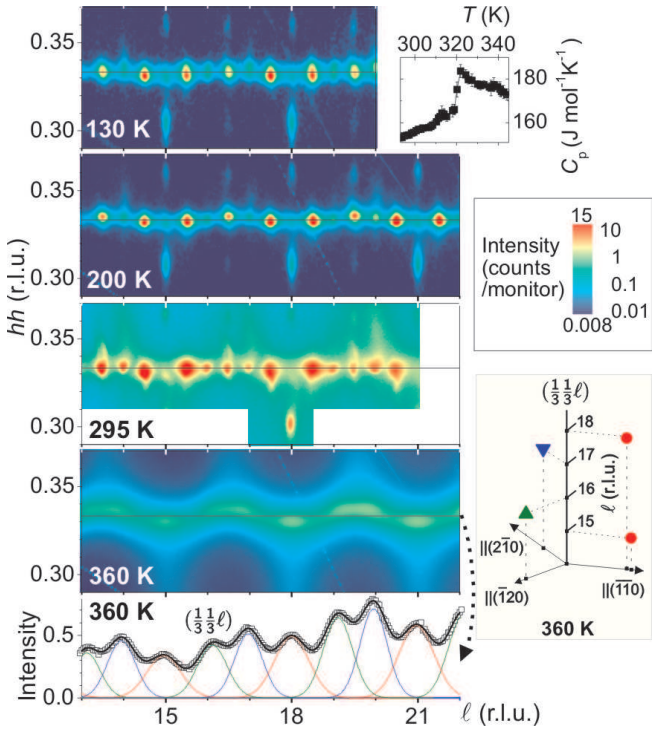}
\caption{ (Color online) Scattered intensity at ($hh\ell$) at
$130$, $200$, $295$, and $360\,{\rm K}$. Lowest panel:
($\frac{1}{3}\frac{1}{3}\ell$) cut at $360\,{\rm K}$ with fit by
Gaussians. Strong structural reflections reach intensities
$>\!10^4\,$counts$/$monitor, strong superstructure reflections
$\sim\! 40$. The higher overall intensity at $295\,{\rm K}$ is due
to wider detector slits being used. Horizontal lines mark
$hh=\frac{1}{3}$. Insets: Specific heat near $320\,{\rm K}$;
sketch of $hk\ell$ positions of maxima in diffuse scattering at
$360\,{\rm K}$, with symbols indicating domains as in Fig.\
\ref{FigCOconfig}(a).} \label{Fighhl}
\end{figure}

We studied single crystals from the same batch as those in
\cite{Christianson08}. X-ray scattering was performed on the (001)
surface of a crystal with a mosaic of $0.02(1)^{\circ}$ at beam
line $6{\rm ID}B$ of the Advanced Photon Source, using $16.2\,{\rm
keV}$ photons. All scattered intensities are normalized to an ion
chamber monitor. Structural Bragg reflections, denoted
$\mathbf{s}$, are consistent with the reported \cite{Isobe90}
rhombohedral $R\overline{3}m$ structure. Specific heat was
measured with commercial equipment.

\begin{figure}[tb]
\includegraphics[width=0.99\linewidth]{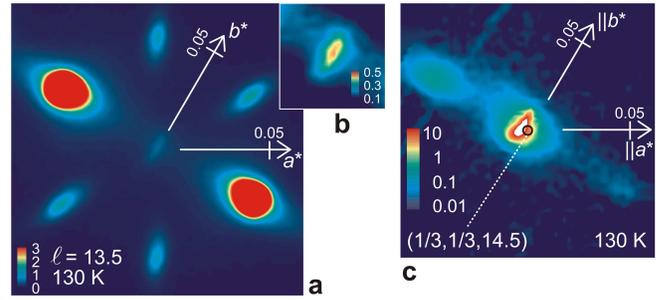}
\caption{ (Color online) Incommensurate satellites. (a) Intensity
at ($hk13.5$) at $130\,{\rm K}$ (the two strongest peaks reach
$\sim\! 19\,$counts/mon). (b) Detail around (0,0,13.5). (c)
Intensity around ($\frac{1}{3}\frac{1}{3}14.5$) (marked by
$\circ$). } \label{Fighk}
\end{figure}

Superstructure reflections (Fig.\ \ref{Fighhl}) appear below a
sharp feature in the specific heat (inset) at $T_{CO} \! \sim
320\,{\rm K}$ indicating the CO transition
\cite{noteMoessLuFe2O4}. We found two sets of strong
superstructure reflections at the same $hk\ell$ positions as
reported in \cite{Yamada0097}, one set near
($\frac{1}{3}\frac{1}{3}\frac{o}{2}$) with $o$ an odd integer, and
satellites to ($00\frac{3o}{2}$) at $\pm$($\tau\tau0$) and (not
previously reported) two symmetry-equivalent directions
[$\tau\!\sim\! 0.03$, see Fig.\ \ref{Fighk}(a)]. As proposed in
\cite{Yamada0097}, the positions of these reflections are
consistent with three CO domains (labelled A, B, and C hereafter)
with the symmetry-equivalent propagation vectors $\mathbf{p}_A =
(\frac{1}{3}+\delta$,$\frac{1}{3}+\delta$,$\frac{3}{2})$,
$\mathbf{p}_B =
(\frac{\overline{2}}{3}-2\delta$,$\frac{1}{3}+\delta$,$\frac{3}{2})$,
and $\mathbf{p}_C =
(\frac{1}{3}+\delta$,$\frac{\overline{2}}{3}-2\delta$,$\frac{3}{2})$.
Here, $\delta\!\sim\! 0.003$ is very small, but $\delta\neq 0$
implies {\em incommensurate} CO. The reflections near
($\frac{1}{3}\frac{1}{3}\frac{o}{2}$) correspond to $\mathbf{s}\pm
\mathbf{p}_.$, and several additional, much weaker, types of
reflections visible in Fig.\ \ref{Fighhl} have positions
consistent with $\mathbf{s}\pm n\mathbf{p}_.$, to be discussed in
detail elsewhere. The reflections shown in Fig.\ \ref{Fighk} may
be considered as higher harmonics as well, likely $\mathbf{s}\pm 9
\mathbf{p}_.$ ($\tau=9\delta$), and are in any case a consequence
of a bimodal charge distribution of the CO configuration obtained
below.

\begin{figure}[t]
\includegraphics[width=0.99\linewidth]{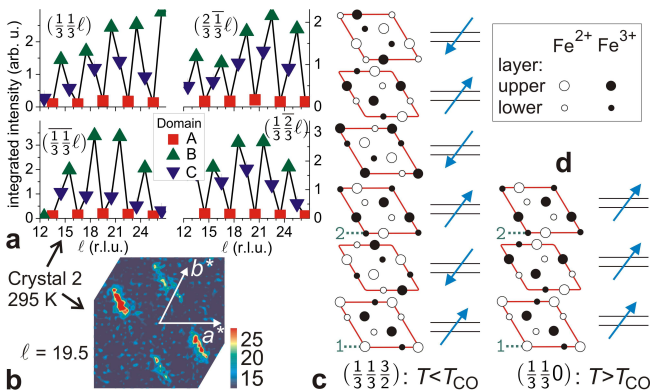}
\caption{ (Color online) Commensurate approximation of the charge
configuration. (a) Integrated intensities of
($\frac{1}{3}\frac{1}{3}\frac{o}{2}$) type reflections.
Reflections associated with different domains are indicated by
different symbols (see text). (b) scattered intensity at
($hk19.5$). (c,d) Charge configuration for domain A for $T<T_{CO}$
(c) and $T>T_{CO}$ (d, short-range). Shown are the Fe ions (see
legend) of the 6 (3) Fe/O double-layers of the
$\sqrt{3}\!\times\!\sqrt{3}\!\times\! 2$ ($\times\! 1$) supercell.
The direction of the local polarization for each double layer is
indicated by an arrow. The two sites of the primitive cell are
numbered (see text).} \label{FigCOconfig}
\end{figure}

To estimate domain populations we collected x-ray diffraction data
on a second crystal \cite{note_twinning}, using a CuK$\alpha$
diffractometer [Fig.\ \ref{FigCOconfig}(a,b)]. Intensities of
superstructure reflections close to
($\frac{1}{3}\frac{1}{3}\frac{o}{2}$) and equivalent directions
are shown in Fig.\ \ref{FigCOconfig}(a). Reflections at positions
$\mathbf{s}\pm \mathbf{p}_A$, $\mathbf{s}\pm \mathbf{p}_B$, and
$\mathbf{s}\pm \mathbf{p}_C$ are indicated by $\blacksquare$,
$\blacktriangle$, and $\blacktriangledown$, respectively. Taking
into account that $R\overline{3}m$ structural reflections occur
only for $-h+k+l=3n$ with $n$ integer, each observed
superstructure reflection is associated with one particular
domain. All measurements are consistent with domain A being almost
unpopulated, and domain B roughly twice more populated than domain
C. The intensity ratios of satellite pairs around ($0,0,19.5$)
[Fig.\ \ref{FigCOconfig}(b)], each pair arising from one domain,
indicate the same domain populations, supporting the hypothesis
that they are resulting from the same CO configuration. The
consistent domain populations obtained from the intensities
confirms the $\mathbf{p}_A$, $\mathbf{p}_B$, $\mathbf{p}_C$,
propagation type proposed above.

To determine the CO configurations corresponding to propagations
$\mathbf{p}_A$, $\mathbf{p}_B$, and $\mathbf{p}_C$, we performed
representation analysis \cite{SARAh,MODY}. Similar to the spin
order determined earlier \cite{Christianson08}, there are two
allowed irreducible CO representations i) and ii), corresponding
to either same or different valence for the $\mathtt{1}$ and
$\mathtt{2}$ Fe sites [Fig.\ \ref{FigCOconfig}(c,d)] of the
primitive cell. For simplicity, our following discussion assumes a
commensurate approximation ($\delta,\tau\rightarrow 0$)
\cite{note_commensurateapprox} in which the propagation vectors
become ($\frac{1}{3}\frac{1}{3}\frac{3}{2}$) and
symmetry-equivalent wavevectors. Both representations contain
sites with two different magnitudes of the Fe valence difference
from average, in clear contrast to the results of M\"{o}ssbauer
spectroscopy, which imply a bimodal valence distribution
\cite{Xu08}. A bimodal charge distribution is only obtained by
adding a uniquely defined ($00\frac{3}{2}$) representation. Case
i) leads to a CO configuration in which the Fe/O double-layers are
not charge neutral. This is physically very unlikely, given the
separation of neighboring double-layers by $\sim\! 6\, {\rm
{\AA}}$. Case ii) leads to a single feasible configuration with
overall neutral double-layers, see Fig.\ \ref{FigCOconfig}(c).
Shown is the configuration for domain A, accommodated in a
$\sqrt{3}\times\sqrt{3}\times 2$ supercell. Note that the CO
lowers \cite{note_atomshift} the crystallographic space group
symmetry to monoclinic $C2/m$. For this CO configuration,
structure factor calculations \cite{note_Ikeda05b} confirm the
observed reflection pattern, including both $\mathbf{s}\pm
\mathbf{p}_.$ and $\mathbf{s}\pm$($00\frac{3}{2}$). The latter, in
the commensurate approximation, correspond to the reflections
shown in Fig.\ \ref{Fighk}(a), again confirming the association of
both these types of reflections with the same CO. As a further
check of the proposed CO [Fig.\ \ref{FigCOconfig}(c)], we used
this solution to re-analyze the $220\,{\rm K}$ neutron data
presented in \cite{Christianson08}. Combining the CO model with
the previous model of spin order, magnetic scattering at {\em
both} ($\frac{1}{3}\frac{1}{3}n$) and
($\frac{1}{3}\frac{1}{3}\frac{o}{2}$) could be refined
successfully \cite{note_neutrons}.

\begin{figure*}[tb]
\includegraphics[width=0.99\linewidth]{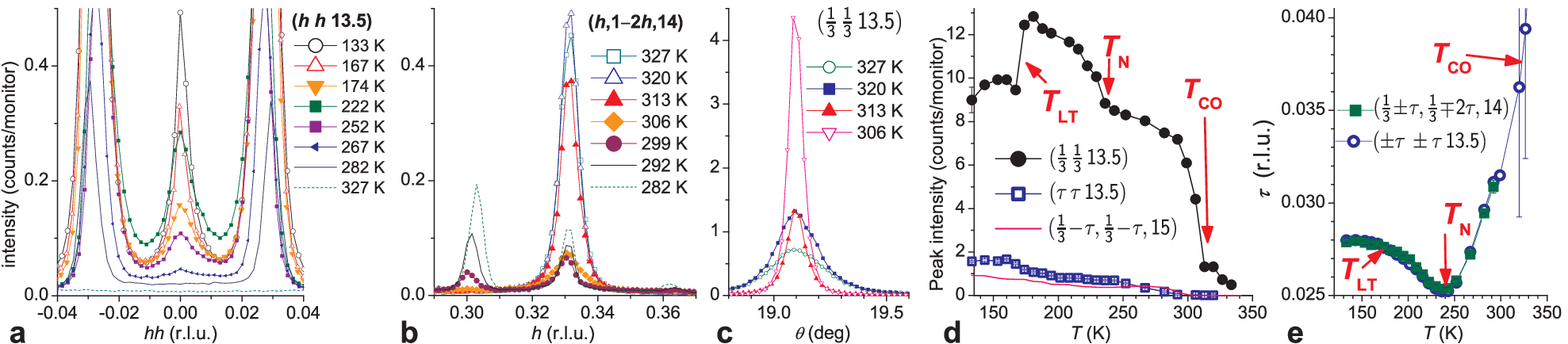}
\caption{ (Color online) $T$ dependence of CO on warming. (a)
($hh13.5$) scans. (b) Scans $\|$($1\overline{2}0$) through
($\frac{1}{3}\frac{1}{3}14$). (c) ($\frac{1}{3}\frac{1}{3}13.5$)
rocking curves. (d) Peak intensities of three reflections vs $T$.
(e) Incommensuration $\tau (T)$. } \label{FigTdep}
\end{figure*}

For the above CO model, the configuration in each individual
double-layer is polar as proposed in \cite{Ikeda05}. However, the
stacking of the polarization of the six double layers of the
supercell [Fig.\ \ref{FigCOconfig}(c)] is {\em anti}ferroelectric
with no net polarization.
To confirm this antiferroelectric configuration, we carried out
DFT calculations for the ferrielectric and antiferroelectric CO
within a $\sqrt{3}\!\times\!\sqrt{3}\!\times\! 2$ cell (in
addition to the $\sqrt{3}\!\times\!\sqrt{3}\!\times\! 1$
calculations of \cite{Xiang07}), that indeed show that the
antiferroelectric CO configuration is more stable than the
ferrielectric one [by $3.2\,{\rm meV}$/fu (formula unit)], and
hence corresponds to the ground state.

Our result raises the question how the remanent polarization
indicated by pyroelectric current measurements \cite{Ikeda05}
could be explained. A simple explanation would be that the samples
are different, since the strong oxygen stoichiometry-dependence of
physical properties is well-known. However, this is unlikely to
apply, for two reasons. First, superstructure reflections
published in the same paper as the polarization results
\cite{Ikeda05} and other papers by the same group (e.g.,
\cite{Yamada0097}), are consistent with
($\frac{1}{3}\frac{1}{3}\frac{3}{2}$) propagation. Our structure
factor calculations indicate that all CO configurations with
non-zero net polarization have ($\frac{1}{3}\frac{1}{3}n$)
reflections of the same order of magnitude as
($\frac{1}{3}\frac{1}{3}\frac{o}{2}$) reflections, whereas no
($\frac{1}{3}\frac{1}{3}n$) reflections are reported in
\cite{Ikeda05,Yamada0097}. Second, preliminary pyroelectric
current measurements with a similar protocol as in \cite{Ikeda05}
suggest a similar remanent polarization as in \cite{Ikeda05}.

Before presenting an alternative explanation, we turn to the
diffuse scattering observed above $T_{CO}$. Heating through
$T_{CO}$, the ($\frac{1}{3}\frac{1}{3}\frac{o}{2}$) and all
satellite reflections weaken [Fig.\ \ref{FigTdep}(c,d)], as
expected, but ($\frac{1}{3}\frac{1}{3}n$) reflections {\em gain}
in intensity [Fig.\ \ref{FigTdep}(b)]. At $360\,{\rm K}$ there is
still considerable diffuse scattering present around the
($\frac{1}{3}\frac{1}{3}\ell$) rod, see Fig.\ \ref{Fighhl}. A cut
in $hh\ell$ along ($\frac{1}{3}\frac{1}{3}\ell$) reveals broad
maxima now at {\em integer} $\ell$ positions, well described by
strongly overlapping Gaussians (lowest panel). The positions of
these diffuse peaks are sketched in an inset to Fig.\
\ref{Fighhl}. They are consistent with
($\frac{1}{3}-\delta'$,$\frac{1}{3}-\delta'$,$0$) propagation and
equivalent. The width of the Gaussians in Fig.\ \ref{Fighhl}
indicates that CO correlations in $c$ direction ($\xi_c$) extend
to 2$-$3 double-layers \cite{note_cut}, i.e.\ they are 3D rather
than 2D, in contrast to \cite{Yamada0097}. Representation analysis
for the above propagation again yielded two irreducible
representations \cite{note_HTrep}, one rejected as unphysical due
to non-charge-neutral double-layers. The remaining representation
in commensurate approximation leads to a CO configuration [Fig.\
\ref{FigCOconfig}(d)] with $\sqrt{3}\!\times\!\sqrt{3}\!\times\!
1$ cell and ferroelectric stacking of the polarization of the
double-layers. Our DFT calculations using the
$\sqrt{3}\!\times\!\sqrt{3}\!\times\! 2$ cell show that this
ferroelectric configuration is less stable than the
antiferroelectric configuration (i.e., the ground state), but only
by $13.4\,{\rm meV}$/fu or $3\%$ of the overall CO gain.

Although with $\xi_c \lesssim c$ the description in terms of CO
configurations may seem somewhat questionable, the result suggests
that the high $T$ correlations favor a ferroelectric arrangement
between neighboring double-layers. It is thus surprising that,
upon long-range charge ordering, the configuration established is
not ferroelectric. Given the small energy-differences between
antiferro$-$ and ferroelectric configurations it seems likely that
cooling with an electric field applied may stabilize long-range
order of the ferroelectric CO configuration, which would explain
the remanent polarization observed (only) after cooling in an
electric field. This idea should be tested by scattering
experiments with electric fields applied in-situ.

In zero electric field, the ground-state has the same {\em basic}
CO configuration with no net polarization at all $T$ below
$T_{CO}$, but specific features vary with $T$. Here, we focus on
the temperature dependence of the incommensuration of the CO. As
in Fe$_2$OBO$_3$ \cite{Fe2OBO3second}, the incommensuration $\tau$
\cite{note_tau} changes with $T$ [Fig.\ \ref{FigTdep}(e)], but in
LuFe$_2$O$_4$ it is much smaller and present at all $T$. In the
fluctuation regime, $\tau$ decreases rapidly upon cooling below
$T_{CO}$, suggesting that the incommensuration may be an important
factor in stabilizing either ferro- or antiferroelectric CO. This
trend is reversed at $T_N\!\sim\! 240\,{\rm K}$, where $\tau \sim
0.025$ has a minimum: The appearance of the seemingly commensurate
\cite{Christianson08} magnetic order results in forcing the CO to
become more incommensurate. In the center of the satellite rings
[Fig.\ \ref{Fighk}(a,b)], weak additional reflections develop upon
cooling through $T_N$ [Fig.\ \ref{FigTdep}(a)]. The emergence of
these additional reflections is not completely understood at
present, but provides an additional indication of coupling between
the CO and the magnetism.

At the low $T$ transition $T_{LT}\!\sim\! 175\,{\rm K}$ reported
in \cite{Christianson08} the width and intensity of these
additional reflections change, and a moderate impact on various
other reflections and $\tau (T)$ is readily visible in Fig.\
\ref{FigTdep}. While a detailed discussion of the impact of the
low $T$ transition on CO and other degrees of freedom is beyond
the scope of this work, we like to point out that i) the basic CO
configuration remains the same upon cooling through $T_{LT}$, and
ii) the CO correlation along $c$ is slightly ($\sim\! 7$\%)
improved compared to $200\,{\rm K}$, suggesting a better
established CO. In contrast, the magnetic order is less
established below $T_{LT}$ as indicated by the broadness of
various magnetic reflections and diffuse magnetic component
reported in \cite{Christianson08}. Thus the ``final compromise''
between CO and magnetism, established below $T_{LT}$ on
approaching the ground state, seems more favorable to the CO.

In conclusion, we show by scattering experiments and DFT
calculations that the charge-ordered polar Fe/O double-layers of
LuFe$_2$O$_4$ have antiferroelectric stacking in the ground state.
Upon long-range ordering, the high-$T$ ferroelectric short-range
correlations revert to this stacking. The incommensurate nature of
the resulting CO is likely relevant for stabilizing either ferro-
or antiferroelectric charge configurations, and reveals the
coupling of the CO to the magnetism, a coupling likely related to
the observed \cite{Subramanian06} large magneto-dielectric effect.

We thank D.\ S.\ Robinson for assistance and J.\ Voigt, H.\ J.\
Xiang, H.\ M.\ Christen, W.\ Schweika, A.\ Kreyssig, Y.\ Janssen,
S.\ Nandi, and A.\ B.\ Harris for discussions. Work at ORNL, NCSU,
and at the MU-CAT sector of APS was supported by the Division of
Materials Sciences and Engineering, Office of Basic Energy
Sciences, US Department of Energy (DE-AC05-00OR22725,
DE-FG02-86ER45259, DE-ACD2-07CH11358, and DE-AC02-06CH11357).

\newcommand{\noopsort}[1]{} \newcommand{\printfirst}[2]{#1}
  \newcommand{\singleletter}[1]{#1} \newcommand{\switchargs}[2]{#2#1}


\begin{thebibliography}{10}
\bibitem{noteAFE3}
We use the term only to indicate the nature of polar arrangements,
not implying
  reversal of polar arrangements in electric fields, though this seems likely
  \cite{Ikeda05} as well.
\bibitem{noteMoessLuFe2O4}
A freezing of charge hopping is confirmed by optical and
M{\"o}ssbauer spectroscopies \cite{Xu08}.
\bibitem{note_commensurateapprox}
This likely describes the CO locally [see, e.g.,
\cite{Fe2OBO3second}] and using the incommensuration in the
representation analysis does not lead to an overall polarization
either.
\bibitem{note_neutrons}
Though with an unexpectedly high magnetic contrast between Fe ions
in the two valence states. Details will be given elsewhere.
\bibitem{note_cut}
This is a lower limit, because the cut does not go through the
peak maxima.
  Cutting at ($\frac{1}{3}\frac{1}{3}\ell$) is the best ``compromise'' (compare
  with the sketch in Fig.\ 1).
\bibitem{note_HTrep}
Here, addition of a (000) component is necessary to obtain a
bimodal charge distribution.
\bibitem{note_tau}
Our discussion of the temperature dependence of the
incommensuration is based on $\tau$ rather than $\delta$ because
the latter is too small to be accurately measured, and
$\tau\propto\delta$.
\end{thebibliography}
\end{document}